\documentclass[prl,twocolumn,showpacs,preprintnumbers,amsmath,amssymb]{revtex4}
\usepackage{graphicx}% Include figure files
\usepackage{dcolumn}% Align table columns on decimal point
\usepackage{bm}% bold math
\usepackage{latexsym,epsfig}

\begin{document}
\preprint{preprint}
\title{Induced supersolidity in a normal and hardcore boson mixture}
\author{$^1$Tapan Mishra\protect\footnote[1]{E-mail: tapan@iiap.res.in},
$^2$Ramesh V. Pai\protect\footnote[2]{E-mail: rvpai@unigoa.ac.in}
and $^1$B. P. Das\protect \footnote[3]{E-mail: das@iiap.res.in}}
\affiliation{ $^1$Indian Institute of
Astrophysics, II Block, Kormangala, Bangalore, 560 034, India. \\
$^2$Department of Physics, Goa University, Taleigao Plateau, Goa 403
206, India. }
\date{\today}

\begin{abstract}
It is well known that the supersolid form of matter can arise in a single
species of cold bosonic atoms in an optical lattice due to long range 
interactions. We present a scenario where a supersolid is induced in one of 
the components of a mixture of two  species bosonic atoms where there is no 
long range interactions. We study a system of normal  and hardcore boson 
mixture with only the former possessing long range interactions. We consider 
three cases: the first where the total density is commensurate and the other 
two where they are incommensurate . By suitable choices of the densities and 
the interaction strengths of the atoms, we predict that the charge density 
wave and the supersolid orders can be induced in the hardcore species as a 
result of the competing interatomic interactions. 
\end{abstract}


\maketitle
The hallmark of the supersolid(SS) phase is the coexistence of the 
superfluid and the charge density wave (CDW); i.e. solid  orders  
~\cite{andreev,leggett}. This phase has not been observed unambiguously 
in experiments. However, in recent years several theoretical predictions of 
this phase have been made in different lattice 
systems~\cite{batrouniprl,scarola,kedar,arun,sengupta}. After the claim of 
the observation of this phase by Kim \textit{et al}~\cite{chan} which later 
could not be verified ~\cite{ref1,ref2,ref3} the search for the supersolid 
phase has become an active area of research.

The pioneering observation of the superfluid (SF) to Mott insulator (MI) 
transition in an optical
lattice using cold bosonic atoms ~\cite{greiner}, 
which had been predicted by Jaksch et. al. \cite{jaksch} based on an earlier 
work by Fisher et. al. \cite{fisher}, has opened up new directions in the 
field of ultracold atoms. The possibility of tuning the interatomic 
interactions in optical lattice makes this system  ideal for obtaining exotic 
phases of ultra cold atoms with long range interactions~\cite{scarola,goral}. 
The observation of BEC in $^{52}Cr$ atoms which have large magnetic dipole 
moments in a trap ~\cite{pfau} in combination with the advancing research in 
optical lattice systems raises the hope of the observation of the supersolid 
phase in the future.

Theoretical studies of the possible existence of supersolids in mixtures of 
bosonic atoms and Bose-Fermi mixtures have attracted much 
attention~\cite{mathey,batrounibf,hofstetter}. Mixtures of ultracold atoms 
are very interesting because of the various competing interactions between 
the atoms leading to many exotic phases. 
It has been shown that for a single species ultracold atoms in a 
one dimensional lattice, the supersolid exists for large onsite as well as 
nearest neighbour interactions when the density of the system is not 
commensurate to the lattice and also larger than half 
filling~\cite{ourpaper,batrouniprl}. In this Letter, we have considered a 
mixture of two species bosonic atoms with one species consisting of normal 
and the other hardcore bosonic atoms. For the latter species, a single lattice 
site can be occupied by no more than one atom. This mixture can therefore be 
considered equivalent to a system of Bose-Fermi(spinless) atoms. 
We assume that the normal bosonic species exhibits long range interactions, 
but the inter atomic interactions in the hardcore species are limited only to 
onsite interactions. The model Hamiltonian for such a system can be written as:
\begin{eqnarray}\label{eq:ham}
\nonumber
H&=&-t^{a}\sum_{<i,j>}(a_{i}^{\dagger}a_{j}+\mbox{H.c.})-t^b\sum_{<i,j>}
(b_{i}^{\dagger}b_{j}+\mbox{H.c.}) \\
\nonumber &&+\frac{U}{2}^{a}\sum_{i}
n_{i}^{a}(n_{i}^{a}-1)
+U^{ab}\sum_{i}n_{i}^{a}n_{i}^{b}\\
&&+V^a\sum_{<i,j>}n_{i}^{a}n_{j}^{a}.
     \end{eqnarray}
Here $a_i$ and $b_i$, respectively, are the  bosonic annihilation operator 
for atoms of $a$ (normal) and $b$ (hardcore) bosons 
localized on site $i$, $n^a_i=a^\dagger_ia_i$ and $n^b_i=b^\dagger_ib_i$ 
represent its number
operators and $t^{a}$ and $t^{b}$ are the hopping amplitudes between the 
nearest 
neighbours $\langle ij \rangle$. $U^a$ ($V^a$) are the on-site 
(nearest neighbour) 
intra-species repulsive interactions for the normal atoms. 
The inter-species ( between normal and hardcore bosons) 
interaction is given by $U^{ab}$. The hopping amplitudes 
($t^a$, $t^b$) and interaction parameters ($U^a$, $V^a$, $U^{ab}$) 
are related to depth of the optical potential, recoil energy and the 
scattering lengths~\cite{jaksch,bloch}. The ratio $U^{ab}/U^{a}$ as well as 
$U^{a}/V^{a}$ can be 
varied over a wide range of values experimentally ~\cite{demler,pfaureview}. 
In this work we consider $t^a=t^b=t$ and set our energy scale by taking $t=1$. 

We identify various ground states phases of the model (\ref{eq:ham}), 
by calculating (i) the single
particle excitation gap $G_L^\alpha$ for species $\alpha=a,b$ defined as the 
difference between the
energies needed to add and remove one atom of species $\alpha$; i.e.,
$G_L^a=E_L(N_a+1,N_b)+E_L(N_a-1,N_b)-2 E_L(N_a,N_b)$, 
$G_L^b=E_L(N_a,N_b+1)+E_L(N_a,N_b-1)-2 E_L(N_a,N_b)$ 
and (ii) the on-site number density defined by
\begin{equation}\label{eq:ni}
    \langle n^{\alpha}_i \rangle=\langle
\psi_{LN_aN_b}| n^{\alpha}_i |\psi_{LN_aN_b}\rangle.
\end{equation}
Here $\alpha$, as mentioned before, is an index representing normal ($a$) or
hardcore ($b$) bosons, with $N_a$ ($N_b$) corresponds to total number of $a$
($b$) bosons in the ground state $|\psi_{LN_aN_b}\rangle$ of a
system of length $L$ with the  ground state energy $E_L(N_a,N_b)$. The former
is used to distinguish the gapless superfluid phase from 
the Mott insulator or the charge density wave phase, both having finite gap in 
their energy spectrum. 
In one dimension the appearance of the SF phase is indicated by 
$G_L^\alpha \rightarrow
0$ for $L\rightarrow \infty$. However, for a finite system $G_L^\alpha$ is
finite, and we must extrapolate to the $L\rightarrow \infty$ limit,
which is best done by the finite size scaling of the gap\cite{pai,paiprl}.
In the critical region
\begin{eqnarray} G_L^\alpha \equiv L^{-1} f(L/\xi^\alpha)
\end{eqnarray}
where $\xi^\alpha$ is the correlation length for species $\alpha$ which 
diverges in the SF
phase. Thus plots of $LG_L^\alpha$ versus the nearest neighbour interaction for 
different values
of $L$ coalesce in the SF phase. On the other hand, when this trend
does not follow, then the system is considered to be in the gapped, 
MI or CDW phase which is further distinguished from each other 
via the CDW order parameter defined as

\begin{equation}
O^{\alpha}_{CDW}(L)=\frac{1}{L}\sum_i \langle\psi_{LN_aN_b}| (|n^{\alpha}_i -
\rho^{\alpha}|)|\psi_{LN_aN_b} \rangle.
\end{equation}
The existence of the solid order in the thermodynamic limit is verified from the
finite value of $O^{\alpha}_{CDW}(L\rightarrow \infty)$.    

We have employed the finite size density matrix renormalization group (FS-DMRG) method with open-boundary conditions to determine the ground state. This method has  proved to be one of the most powerful techniques for studying 1D systems ~\cite{white,paiprl,pai,dmrgreview}. 
For the normal species, we have taken the maximum occupation per site as four ($4$). 
We allow up to $128$ states in the density matrix of the
left and right blocks in each iteration of the FS-DMRG calculations. 
The weights of the states neglected in the density matrix of the left and 
right blocks are less than $10^{-6}$.

The charge density wave phase in bose systems is possible when the 
density of bosons are commensurate with the underlying lattice.
For example, the earlier studies of the  
one-dimensional single species extended Bose-Hubbard model have shown 
the existence of the CDW phase for 
$\rho=1/2, 1$~\cite{pai,batrouniprl,ourpaper,kuhner}. Later this study was
extended 
in the case of two species extended Bose-Hubbard model, 
where  the solid order is achieved for
$\rho^a=\rho^b=1/2$ by suitably varying the strengths of the 
nearest neighbour interactions~\cite{sahoo}. 
Supersolid phase is then possible only moving away from these commensurate 
densities. In a recent study on a Bose-Fermi mixture with different hopping amplitudes, the supersolid 
phase has been predicted without the nearest neighbour interactions by 
doping the bosonic species~\cite{batrounibf}. In order to achieve the 
supersolid phase in these systems, the inter atomic interactions have been 
carefully controlled  such that the added bosons do not destroy the CDW phase 
by occupying sites that are already occupied.  

The recent study of a two 
species Bose mixture in a one dimensional lattice shows that phase separation 
occurs if the ratio $U^{ab}/U^{a}$ is larger than unity~\cite{mishraphasesep}. 
In order to avoid this condition, we consider $U^a=U^{ab}=U$ and study the 
effect of $V^a$ on the ground state of model (\ref{eq:ham})
for three possible combinations of densities;(i) 
$\rho^a=\rho^b=1/2$,(ii) $\rho^a=1/2$ and $\rho^b=1/4$ and (iii)$\rho^a=3/4$ 
and $\rho^b=1/2$. In the first case, the total density of the system is 
commensurate, but in other two cases it is  not. In all the above three 
cases we have taken $U=6$,  which is very large compared to the nearest 
neighbour tunneling amplitude $t=1$. 

\begin{figure}[tbp]
  \centering
  \epsfig{file=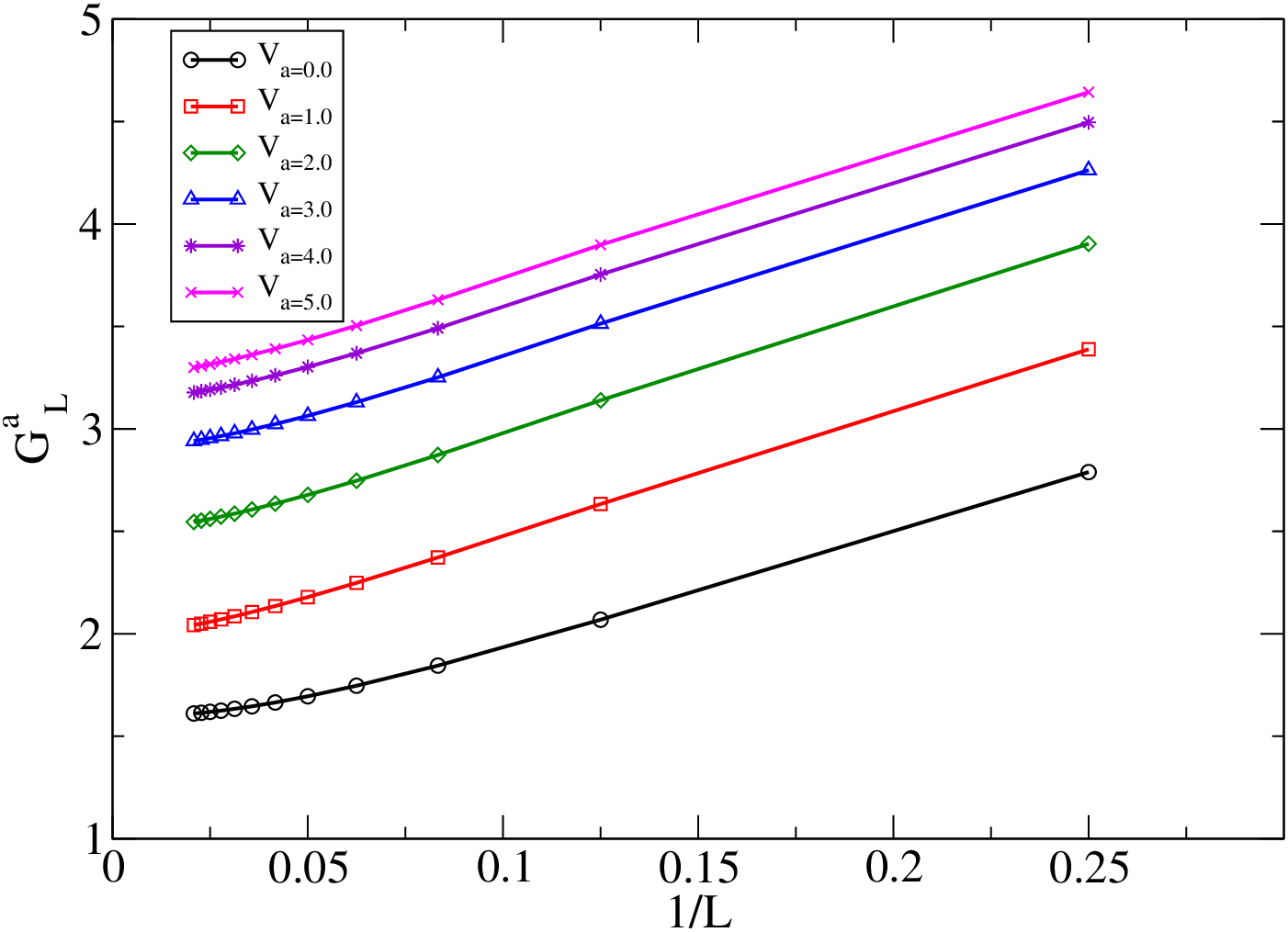,width=8cm,height=6cm}
  \caption{(Color Online)Single particle energy gap $G^a_L$ for normal bosons
as a function of $1/L$ for
    different $V^a$ showing finite gap 
in the thermodynamic limit.}
     \label{fig:fig1}
\end{figure}

\begin{figure}[tbp]
  \centering
\includegraphics[width = 3.0in, height=6cm, angle = 0, clip = true]
{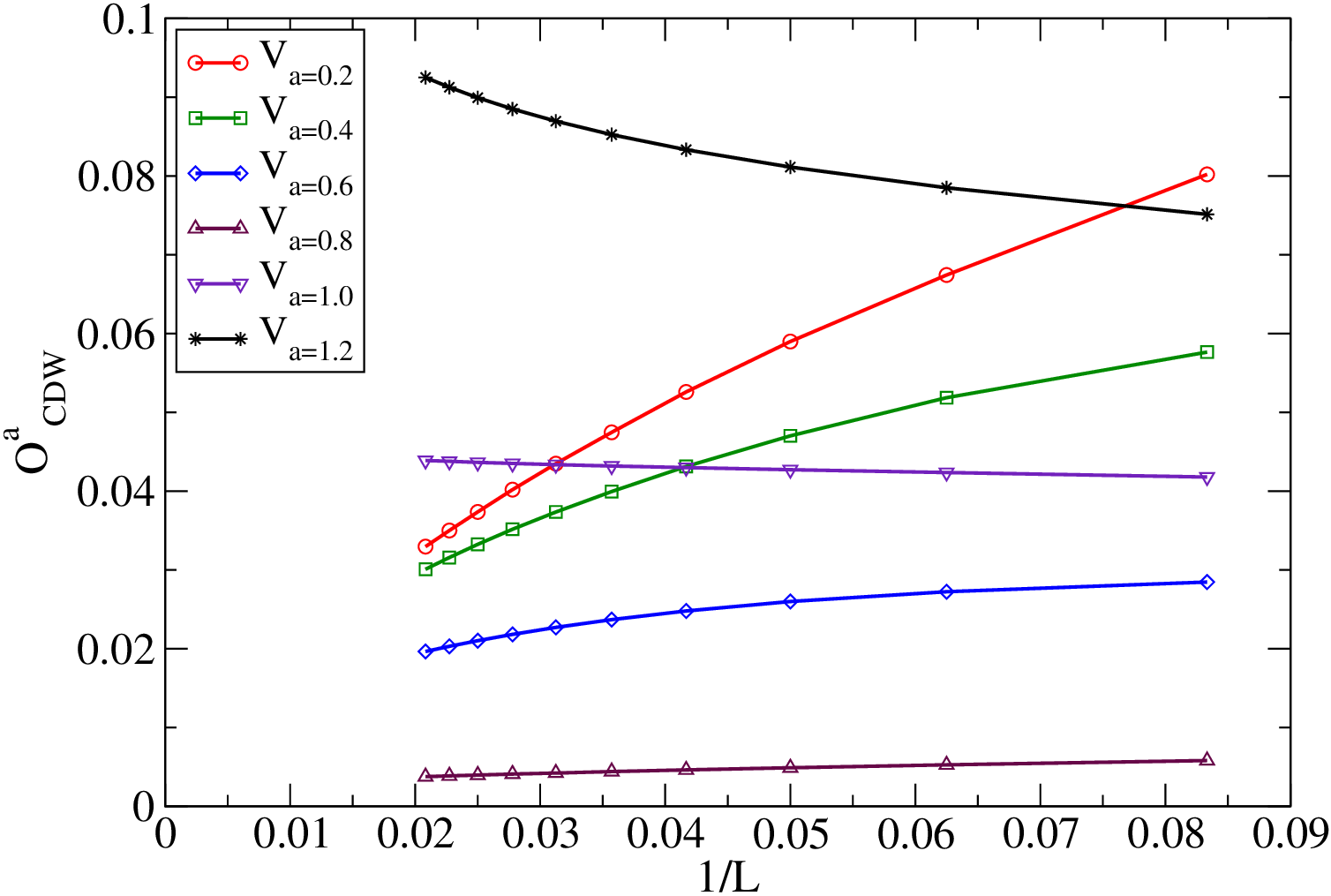}
  \caption{(Color online)
$O^a_{CDW}(L)$ versus $1/L$ for different $V^a$ showing the onset of CDW order 
for $V^a > V^a_C\approx1.0$.  }
     \label{fig:fig2}
\end{figure}

$\rho^a=\rho^b=1/2$ :- In this case the total density of bosons 
$\rho=\rho^a+\rho^b=1$. For  $V^a=0$, the system is in the MI phase
because the onsite intra species interactions, $U^a=6$, 
$U^b=\infty$ and the inter species interaction, $U^{ab}=6$ 
are all greater than $U_C\approx 3.4$, the critical strength of the 
on-site interaction for the SF-MI transition in the one-dimensional 
Bose-Hubbard model~\cite{pai,kuhner,mishraphasesep}. The system continues to remain gapped as $V^a$ increases. The gap corresponding to lattice size L for species a,
 $G_L^a$ is plotted for different values of $V^a$ in 
Fig.\ref{fig:fig1}. This figure clearly shows that  $G_{L\rightarrow \infty}^a$
is finite for all the values of $V^a$ that we have considered. However, the  
gapped phase at higher
$V^a$ is not a MI but a CDW, since $O_{CDW}^a(L\rightarrow \infty)$ 
is finite for $V^a>V^a_{C}\sim 1.0$. 
In Fig.\ref{fig:fig2} we have plotted 
$O_{CDW}^a(L)$ versus $1/L$ for different values of $V^a$. 
The order parameter $O_{CDW}^a(L)$ goes to zero for small values of $V^a$ 
and branches out for higher values indicating the onset of the CDW phase.
$O_{CDW}^b(L)$ also exhibits a similar behaviour indicating that both the normal and
hard-core bosons undergo a MI to CDW transition.   
The dependence of this transition on $V^a$ for the normal bosons is expected
on the basis of an earlier work~\cite{kuhner}. However, it was not obvious that 
the hard-core bosons would also undergo
a similar MI to CDW transition as they lack long range interactions to
exhibit density oscillations.

\begin{figure}[htbp]
  \centering
\includegraphics[width = 3.0in, height=7cm, angle = 0, clip = true]
{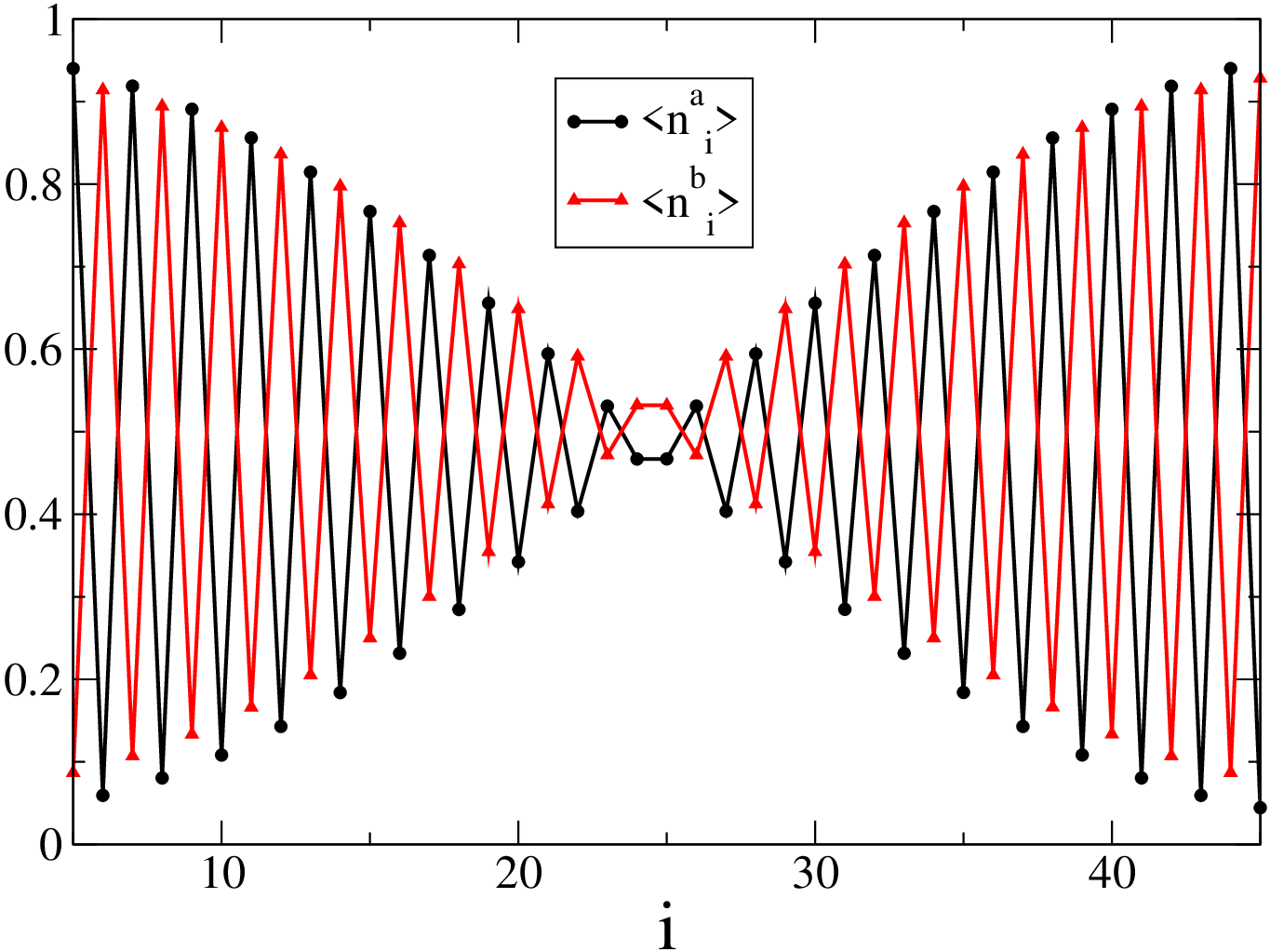}
\caption{(Color online) The onsite number density for $a$ and $b$ bosons are 
plotted against the site index $i$ taking $V^a=5$ 
for $\rho^a=1/2$, $\rho^b=1/2$. 
The black curve corresponds to $<n^a_i>$ and the red corresponds to $<n^b_i>$. 
The node at the center is due to open boundary condition.}
	\label{fig:fig3}
\end{figure}

The physical scenario when $\rho^a=\rho^b=1/2$ is the following: when $V^a=0$, 
due to the strong repulsion between the bosons, the system is in the MI phase
with both normal and hard-core bosons uniformly distributed through out the 
lattice giving the average density of the total number of bosons at every site  equal 
to one. As $V^a$ increases, there is competition between the 
interactions, $U^{ab}$, $U^a$ and $V^a$ and hence an atom of species $a$
(normal bosons)  
cannot occupy the sites next to another atom of the same species, 
thereby forming a CDW phase. In addition, the atoms of the hardcore species $b$ 
cannot occupy a site where there is either a hardcore boson or a normal boson
because of the strong repulsive onsite interaction, 
$U^{ab}$. These physical conditions give rise to the intermingled CDW phase 
where the 
nearest neighbour sites are occupied by atoms of different species as shown 
in (Fig.\ref{fig:fig3}). 
It is interesting to note that the presence of $V^a$ is sufficient to induce 
the solid order in the hardcore species in spite of the absence of  any long range 
interaction between them. 
This type of induction of the solid order makes the other 
combinations of densities presented below very interesting. 

\begin{figure}[htbp]
  \centering
\includegraphics[width = 3.0in, height=7cm, angle = 0, clip = true]
{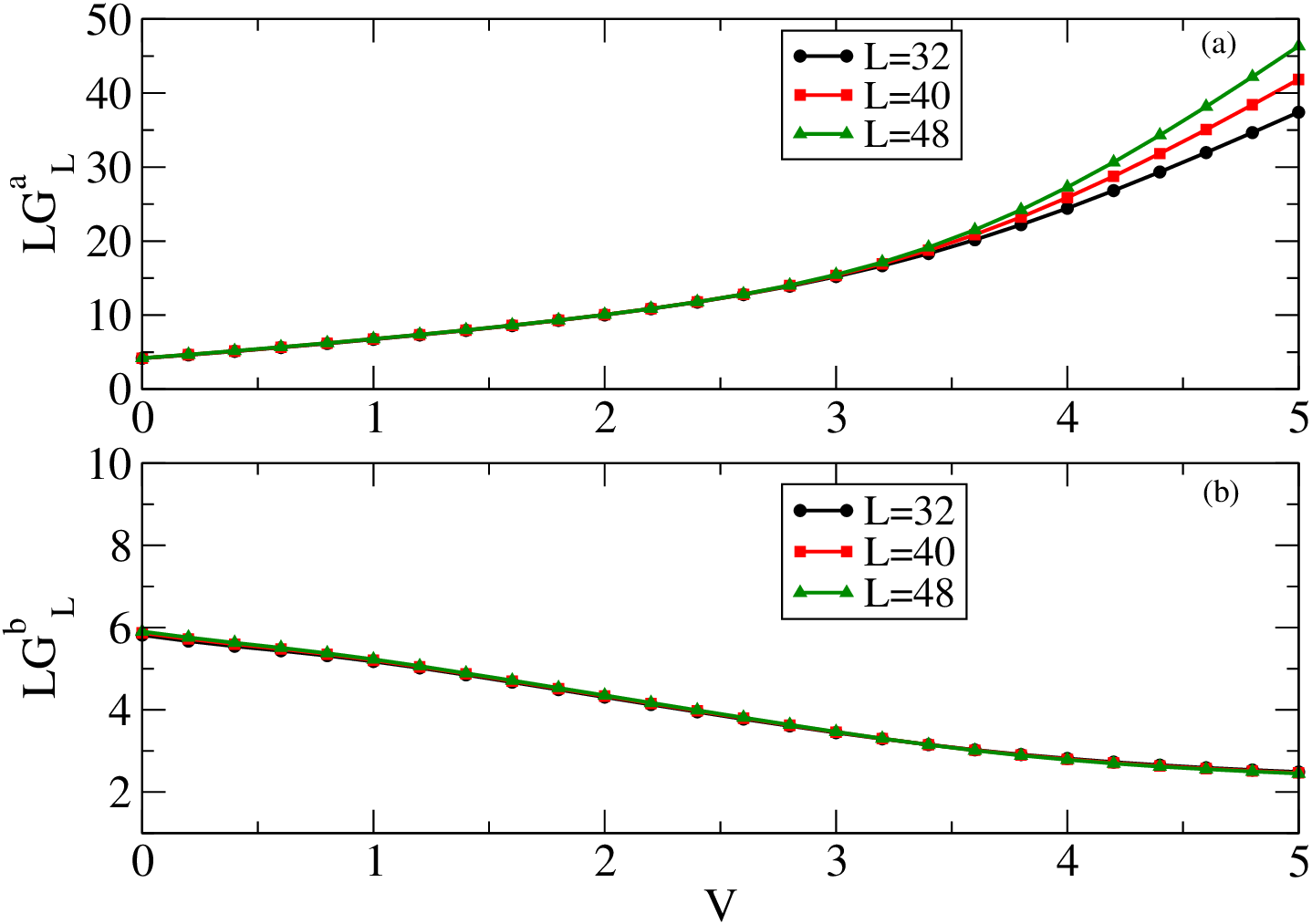}
      \caption{(Color online)Finite size scaling of gap $LG_L^\alpha$ 
is plotted as a function of $V^a$ for
different system sizes for $\rho^a=1/2$ and $\rho^b = 1/4$. 
(a)The coalescence of the
different curves for $V^a \leqslant 3.0$ for the normal species shows 
the transition from the gapless SF phase to a gapped phase. 
(b) The coalescence of the different curves for all the values of $V^a$ 
for the hardcore species indicates the 
existence of the SF phase.}
     \label{fig:fig4}
\end{figure}

\begin{figure}[htbp]
  \centering
\includegraphics[width = 3.0in, height=7cm, angle = 0, clip = true]
{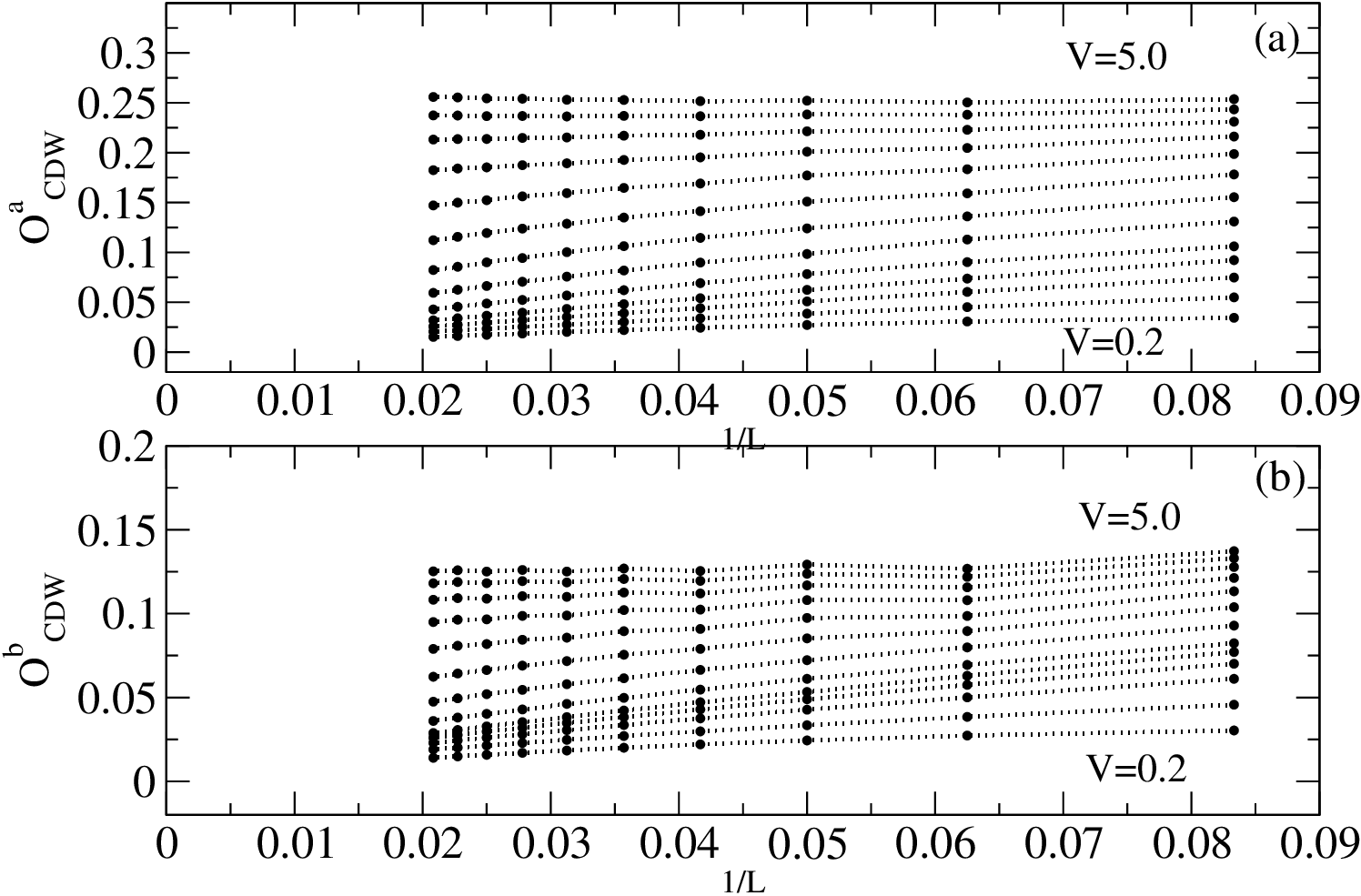}
	\caption{$O_{CDW}(L)$ versus $1/L$ for (a) normal and (b) 
hardcore species for $V^a$ ranging from $0.2$ to $5.0$ in steps of $0.4$ 
is shown for $\rho^a=1/2$ and $\rho^b=1/4$ . It is clear from the scaling that 
the $O_{CDW}^\alpha$ becomes finite for $V^a\sim 3.0$ for both the species 
indicating the transition to the CDW phase.}
	\label{fig:fig5}
\end{figure}

$\rho^a=1/2$ and $\rho^b=1/4$ :- In this case the total boson density 
$\rho=3/4$, and it is not commensurate with the lattice. In the present problem, we have not considered long range interactions beyond the nearest neighbour, the commensurate densities are therefore intergers or half integers. In a normal two species bosonic mixture with 
incommensurate density (e.g.$\rho=3/4$), there is no transition from a 
SF to a gapped phase ~\cite{mishraphasesep}. However, such a transition does occur in a normal - hardcore boson mixture described by the model (\ref{eq:ham}). 
The finite size scaling of the gap $LG^\alpha_L$ that is obtained from our FS-DMRG calculation shows,  a transition from the gapless SF phase to a gapped phase for the normal bosonic species  in  Fig.\ref{fig:fig4}. The critical value of $V^a$ is $V^a_{C}\sim 3.0$. However, the hardcore species remains in the SF phase 
showing no gap in the excitation spectrum.
The calculation of the CDW order parameters given in Fig. \ref{fig:fig5}
 for both the normal and the hardcore bosons 
show a finite $O_{CDW}^\alpha$ in the limit $L\rightarrow \infty$ 
for $V^a > V^a_C\sim 3.0$. Thus the gapped phase of the normal bosonic 
species is identified to be the 
CDW phase while the gapless phase of hardcore bosons is  a supersolid since
both sperfluid and CDW coexist. The CDW oscillations are similar to the case
of $\rho_a=\rho_b=1/2$ as given in Fig.\ref{fig:fig3}.

\begin{figure}[htbp]
  \centering
\includegraphics[width = 3.0in, height=7cm, angle = 0, clip = true]
{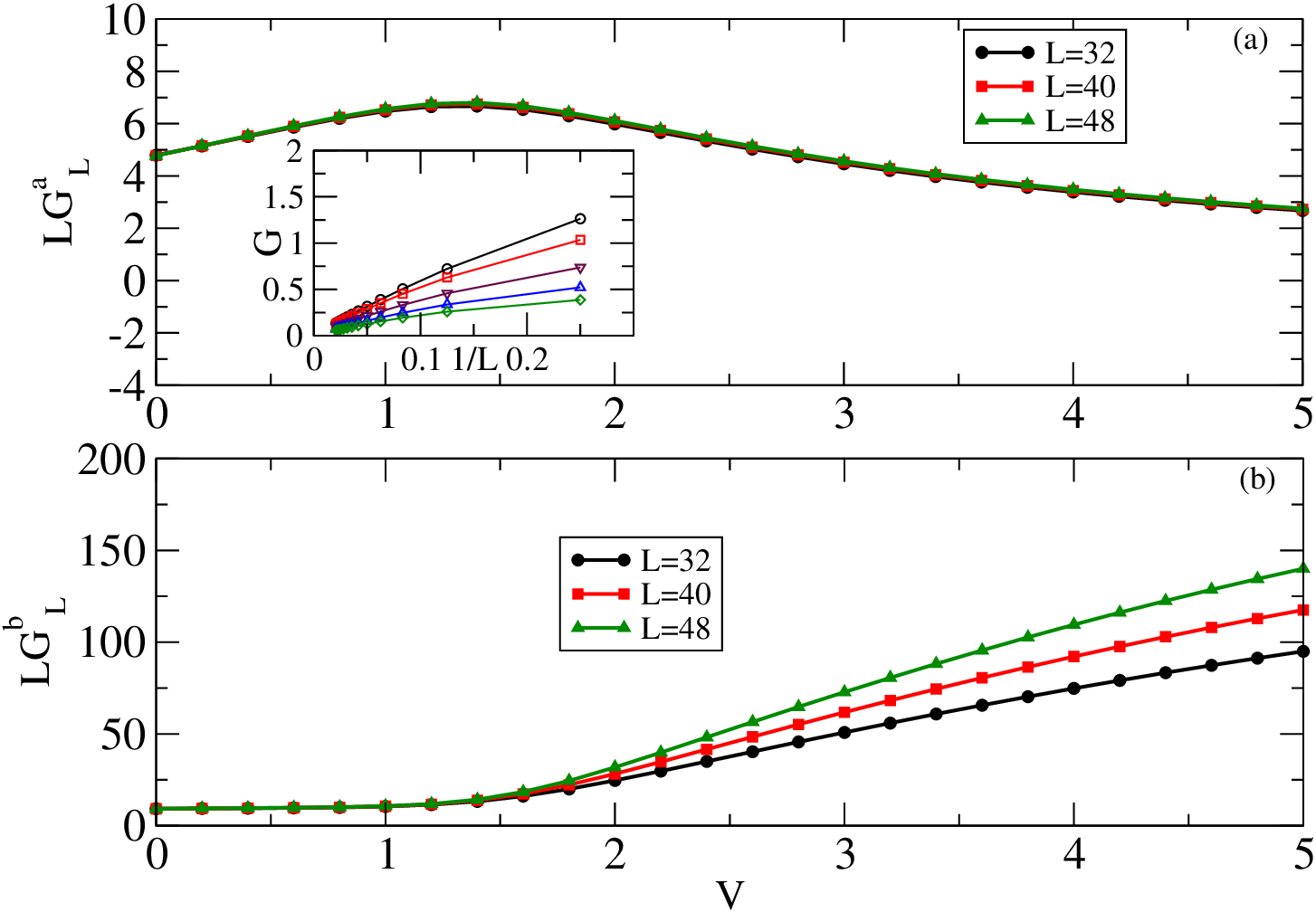}
	\caption{(Color online)Scaling of gap $LG_L$ is plotted as a function 
of $V^a$ for
different system sizes for $\rho^a=3/4$ and $\rho^b = 1/2$. 
(a)The coalescence of the
different curves for all the values of $V^a$ for the normal species 
shows that the system remains in the gapless SF phase.(Inset)Shows the 
$G^a_L$ plotted against $1/L$ for different $V^a$ ranging from $1$ to $5$ 
in steps of $1$ which shows that the gap vanishes in the thermodynamic 
limit as the curves extrapolate to zero as $L\rightarrow \infty$(b) The 
coalescence of the different curves for $V^a \leqslant 1.2$ for the hardcore 
species indicates the transition from the SF to the CDW phase.}
	\label{fig:fig6}
\end{figure}

\begin{figure}[htbp]
  \centering
\includegraphics[width = 3.0in, height=7cm, angle = 0, clip = true]
{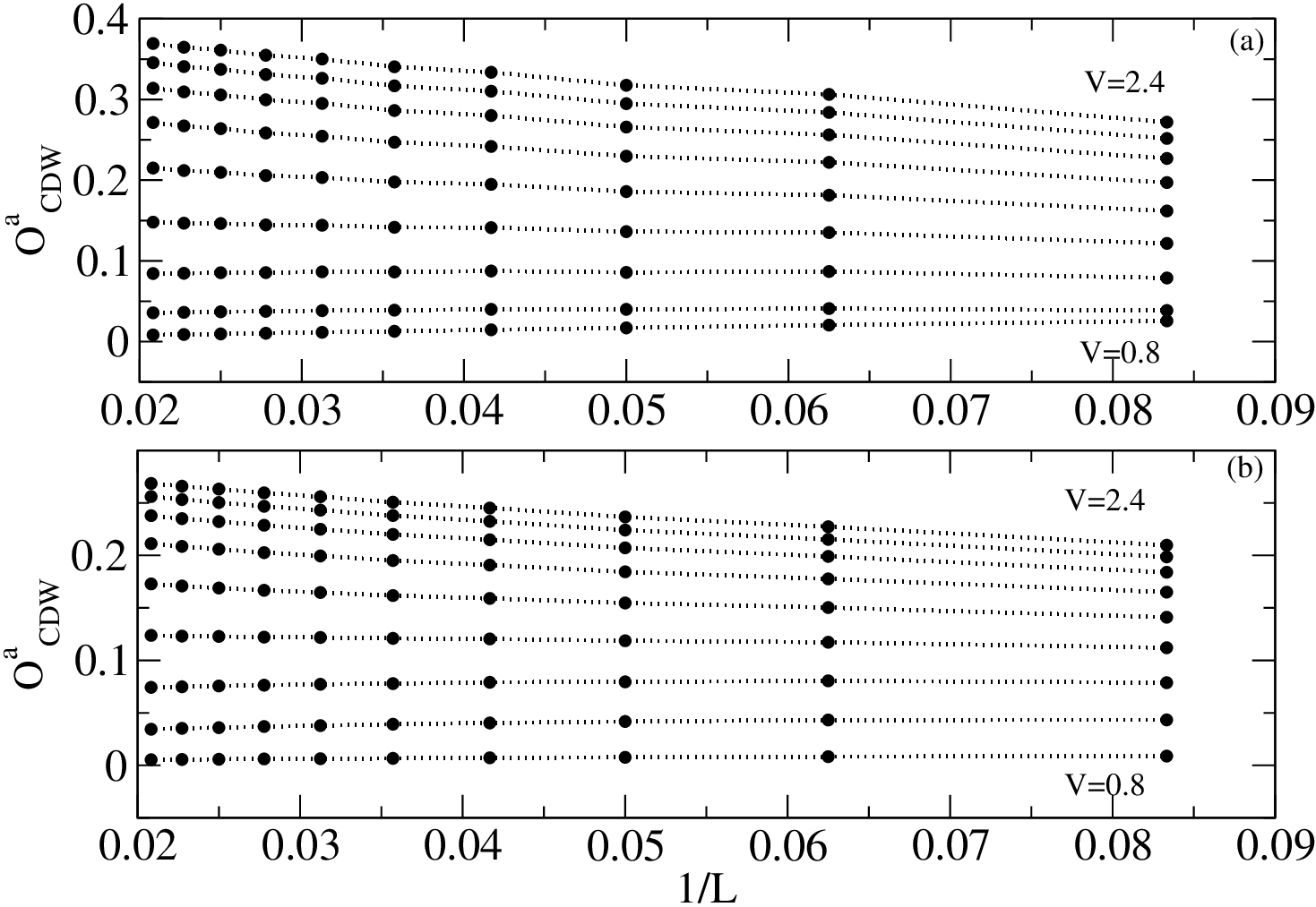}
	\caption{Finite size scaling of $O_{CDW}$ for (a) softcore and 
(b) hardcore species for $V^a$ ranging from $0.8$ to $2.4$ in steps of 
$0.2$ is shown for $\rho^a=3/4$ and $\rho^b=1/2$ . It is clear from the 
scaling that the $O_{CDW}$ becomes finite for $V^a\sim 1.2$ for both the 
species indicating the transition to CDW phase.}
	\label{fig:fig7}
\end{figure}

$\rho^a=3/4$ and $\rho^b=1/2$ :- The phase transitions that we have 
obtained for this case for the two species are in the reverse order 
as that of case (ii).  The normal species 
shows a 
transition from the SF to the supersolid phase not showing any gap in the 
excitation spectrum (see Fig.\ref{fig:fig6}) and a finite CDW order 
parameter at $V^a_{c}\sim 1.2$ (see Fig.\ref{fig:fig7}). However, the 
hardcore species makes a transition from the SF to the CDW phase at the same 
critical point $V^a_c\sim 1.2$. 

In recent experiments on dipolar atoms, it has been shown that the the ratio between the on-site interaction and the nearest neighbour interaction can be controlled by Feshbach resonance~\cite{pfaureview}. In the present work, we take a fixed on-site interaction and vary the nearest neighbour interaction, which is equivalent to the experimental situation where the ratio is varied. Our results suggest a way to obtain the supersolid phase, where we have restricted the control parameter only to the {\it s-wave} scattering length of one species which induces supersolidity in the other species. 

\textit{Conclusion}:- We have considered a system of normal and 
hardcore bosonic mixture with the normal species possessing long range 
interactions. By taking three different sets of densities of both the 
species, we have investigated the conditions that give rise to the 
supersolid phase in either or both the species. The main findings of this 
work is that by suitably tuning the nearest neighbour interaction 
strength $V^a$, the solid order can be stabilized in the normal bosonic 
species and it can also be induced in the hardcore species as a 
result of the competition between the $U^a$, $U^{ab}$ and $V^a$. This 
induction of the solid order can lead to both the species being in the 
CDW and more interestingly one species in the CDW phase and the other in the 
supersolid phase depending on the choice of densities. By keeping the 
onsite repulsion for the normal species $U^a=6$ and varying the nearest 
neighbour interaction strength of the hardcore species $V^a$, we obtain 
different interesting quantum phases which are listed in the table shown below.

\begin{table}[htbp]
\caption{Different phases}
\centering
\begin{tabular}{c c c c}
\hline\hline
$\rho_a$ & $\rho_b$ & softcore & hardocre \\[2ex]
\hline
1/2 & 1/2 & MI-CDW & MI-CDW \\ 
1/2  &  1/4  &  SF-CDW  &  SF-SS \\ 
3/4 & 1/2 & SF-SS & SF-CDW \\ [2ex]
\hline
\end{tabular}
\label{table:sumary}
\end{table}
If the individual density of a species is commensurate then there exists a SF-CDW or MI-CDW 
transition where as if it is incommensurate then the transition is from 
SF to SS. It is interesting to note that for incommensurate fillings the transition from the SF phase to the CDW or SS phases occur at the same critical value $V^a_{C}$ for both the species. 

\section{ACKNOWLEGMENT}
R. V. P. acknowleges financial support from CSIR, India 
Grant No. 03(1107)/08/EMR-II and DST, India Grant No. SR/S2/CMP-0014/2007.

\end{document}